\documentclass[namedreferences]{SolarPhysics}
\usepackage[optionalrh,solaenum]{spr-sola-addons} % For Solar Physics
\usepackage{epsfig}                     % For eps figures, old commands
\usepackage{graphicx}                    % For eps figures, newer & more powerfull
\usepackage{color}                       % For color text: \color command
\usepackage{url}                         % For breaking URLs easily trough lines
\begin{document}

\begin{article}
%\thesaurus{02.16.1; 06.03.2; 06.06.3; 06.13.1; 06.24.1}
%\documentclass[namedreferences]{SolarPhysics}
%\usepackage[optionalrh,solaenum]{spr-sola-addons} % For Solar Physics
%\usepackage{epsfig}          % For eps figures, old commands
%\usepackage{graphicx}        % For eps figures, newer and more powerfull
%\usepackage{courier}         % Change the \texttt command to courier style
%\usepackage{natbib}         % For citations: redefine \cite commands
%\usepackage{amssymb}        % useful mathematical symbols
%\usepackage{color}           % For color text: \color command
%\usepackage{url}             % For breaking URLs easily trough lines
%\def\UrlFont{\sf}            % define the fonts for the URLs

\begin{opening}

\title{Location of Decimetric Pulsations in Solar Flares}

\author{Arnold O.~\surname{Benz}$^{1,2}$\sep
        Marina~\surname{Battaglia}$^{1,3}$\sep
        Nicole~\surname{Vilmer}$^4$
        }
\runningauthor{Arnold O. Benz et al.}
\runningtitle{Decimetric Pulsations}

\institute{$^1$ Institute of Astronomy, ETH Zurich, 8093 Z\"urich, Switzerland
    e-mail: \url{benz@phys.ethz.ch}\\
    $^2$ Institute for 4D Technologies, FHNW, 5210 Windisch, Switzerland\\
    $^3$ School of Physics and Astronomy, University of Glasgow, Glasgow, G12 8QQ, UK\\
    $^4$ LESIA, Observatoire de Paris, CNRS, UPMC, Universit{\'e} Paris-Diderot, 5 place Jules Janssen 92195 Meudon Cedex France\\
    }

\date{Received: xxx; accepted: xxx}
%\maketitle
%\subsubsection{Abstract}
\begin{abstract}
This work investigates the spatial relation between coronal X-ray sources and coherent radio emissions, both generally thought to be signatures of particle acceleration. Two limb events were selected during which the radio emission was well correlated in time with hard X-rays. The radio emissions were of the type of decimetric pulsations as determined from the spectrogram observed by Phoenix-2 of ETH Zurich. The radio positions were measured from observations with the Nan\c{c}ay Radioheliograph between 236 and 432 MHz and compared to the position of the coronal X-ray source imaged with RHESSI. The radio pulsations originated at least 30 - 240 Mm above the coronal hard X-ray source. The altitude of the radio emission increases generally with lower frequency. The average positions at different frequencies are on a line pointing approximately to the coronal hard X-ray source. Thus, the pulsations cannot be caused by electrons trapped in the flare loops, but are consistent with emission from a current sheet above the coronal source.

\end{abstract}
%\keywords{Sun: corona, flares, radio events, particle acceleration, energy release}
\end{opening}
%------------------------------------------------------------------------------
\section{Introduction}
Does the very process of particle acceleration in solar flares directly produce coherent radio emission? Waves such as proposed for current instabilities and stochastic acceleration may couple into radio waves (Benz and Wentzel 1981; Karlick{\'y} and Barta 2005; Li and Fleishman 2009). Alternatively, instabilities produced by non-stochastically accelerated electrons having a non-Maxwellian velocity distribution may be observed by their coherent radio emission within the acceleration site.

In the impulsive phase of flares, very intense radio emissions are often observed. At decimeter wavelengths the peak flux density can reach 10$^6$ solar flux units (10$^{10}$ Jy, Benz 2009). Coherent emissions are most intense and to be distinguished from incoherent radio emissions, such as synchrotron or thermal radiation, not studied here (e.g. Nindos et al. 2008). There are several types of coherent emission processes. Most intense and very frequent are broadband radiations pulsating at irregular intervals on the time scale of one second. As the emission appears to be a non-thermal process, such decimetric pulsations are generally assumed to be tracers of non-thermal electrons and their acceleration.

Radio 'outbursts' at decimeter wavelength during flares have been detected very early with single frequency instruments (Lehany and Yabsley 1948). These authors even remarked the coincidence with an ionospheric radio fadeout, caused by solar flare soft X-ray emission. Covington (1951) noticed occasional high circular polarization of bursts at 10.7 cm. De Feiter, Fokker, and Roosen (1959) reported an association of 545 MHz bursts with H$\alpha$ flares increasing with flare importance. First spectral observations in the decimeter range were reported to consist to 90\% of a 'generalized class of fast-drift bursts' (Young et al. 1961; Kundu et al. 1961). Some of these appear to be the decimetric continuation of meterwave type III bursts, but many others had a different nature, occurring in large groups and showing an 'immeasurably' high drift in frequency. Decimetric emission has also been named 'flare associated continuum' (Pick 1986). Thompson and Maxwell (1962) refer to them as pulsating structures. This notation or simply 'pulsations' has established itself in the literature and will be used here.

The regularity of the pulsations has been noted early. Gotwols (1972) reports a quasi-periodic pulsation over most of the observed band from 600 - 1000 MHz. Remarkably regular pulses at 1.0 s period from 300 - 350 MHz were the basis of the theory of Roberts, Edwin, and Benz (1984) on magnetohydrodynamic oscillations in the corona. In their catalogues, G\"udel and Benz (1988) and Isliker and Benz (1994) characterize pulsations in the decimeter range between 'almost periodic' and 'irregular' with pulse separations of 0.1 to 1 second. Some complex cases may be the superposition of several pulsations with different periods (M{\'e}sz{\'a}rosov{\'a},  Stepanov, and Yurovsky  2011). Ultra-rapid pulsations have been reported by Magdalenic et al. (2003) and Fleishman,  Stepanov, and Yurovsky (1994). Pulsations have better defined upper and lower bounds in frequency than type III bursts and higher drift rates by a factor of 3 on average (Aschwanden and Benz 1986). Contrary to type III bursts, pulsations are highly circularly polarized, except when occurring near the limb  (Aschwanden 1986, 2006).

The frequency range of pulsations extends from meter to centimeter wavelengths, but their character changes. At meter waves, McLean et al. (1971) observed about 50 strikingly regular pulses with periods increasing from 2.5 s to 2.7 s in time. Pulsations above about 300 MHz are less regular. The highest frequency pulsations reported extend beyond 4 GHz (Saint-Hilaire and Benz 2003; Tan et al. 2010) and consist of irregular pulses.

The emission process of pulsations is unclear. It is often associated with some velocity space instability of non-thermal electrons, such as a loss-cone instability or plasma emission by beams (Benz 1980; Fleishman, Stepanov, and Yurovsky 1994). The frequency of such emissions is at the plasma frequency $\nu_p$, the electron gyrofrequency $\nu_e$, the upper hybrid frequency $(\nu_p^2 + \nu_e^2)^{1/2}$, or at twice these characteristic frequencies (review by Benz 2002).  For the driver of such a pulsating instability, electrons trapped in flare loops have been evoked initially (Aschwanden and Benz 1988). More recently, Kliem, Karlick{\'y}, and Benz (2000) proposed that pulsations originate in large-scale current sheets instable to tearing mode reconnection leading to magnetic islands. Their coalescence into a continuously growing plasmoid causes quasi-periodic acceleration. Fleishman, Bastian, and Gary (2008) find observational evidence for this hypothesis. It is also supported by the observed relation to plasmoid ejection observed in X-rays (Khan et al. 2002; Barta, Karlick{\'y}, and Zemlicka 2008; Aurass, Landini, and Poletto 2009). However, Karlick{\'y}, Zlobec, and Meszarosova (2010) conclude that MHD oscillations are more likely in the case of observed sub-second periods. As possible signatures of coronal MHD oscillations, decimetric pulsations have received recent attention (Nakariakov et al. 2010). An alternative origin for short pulses is a periodic self-organizing system of loss-cone instability and particle escape described by Lotka-Volterra coupled equations (Aschwanden and Benz 1988).

Bremsstrahlung of non-thermal flare electrons is readily observed in hard X-rays (HXR) and is therefore also a signature of particle acceleration. HXR observed in coronal sources (review by Krucker et al. 2008) are generally assumed to originate close to the acceleration site. Thus, originating both from super-thermal electrons, the question of the relation of decimetric emissions to HXR emission is immediate. Radio sources at 450 MHz and lower are often observed far from the coronal source. This is well known for type III bursts (e.g. Vilmer et al. 2002). Battaglia and Benz (2009) reported large spatial separations between decimetric spike emissions and the associated, but not time correlated HXR coronal source. On the other hand, Saint-Hilaire and Benz (2003) observed that pulsations occurring nearly simultaneously with the HXR peaks are located close to the HXR source ($\approx$ 10$''$). Their flare was at 0.8 solar radii from the center of the disk, but projection effects cannot be excluded. Later in the flare, pulsations became stronger, but drifted to lower frequency and occurred far from the HXR source.

Decimetric pulsations are well associated with HXR. In a survey between 150 MHz and 4000 MHz, Benz et al. (2005) find 160 pulsations in 201 HXR flares observed by the {\it Ramaty High Energy Solar Spectroscopic Imager} (RHESSI; Lin et al. 2002). This percentage of 80\% is higher than for other decimetric emissions such as spikes (14\% ) and fine structures like intermediate drift bursts and parallel drifting bands (1.5\%). One third of the pulsations are correlated in time with HXR flux in some detail (Dabrowski and Benz 2009).

Here we compare for the first time imaging observations of pulsations and HXR events that have previously been found to correlate in time. We report on the location of decimetric pulsations relative to the coronal HXR source in two well observed limb events. The ultimate question is: What can decimetric pulsations tell us about flare energy release?

\section{Observations, Selection, and Data Analysis}

Solar radio burst types are classified by their characteristics in spectrum and time.  Data from the Phoenix-2 spectrometer (Messmer, Benz, and Monstein 1999) were used. It operated near Bleien, Switzerland (8$^\circ 6' 44''$ E, 47$^\circ 20' 26''$ N), during the time of the selected events. Its broad spectrograms were essential to identify pulsations. A parabolic dish with a diameter of 7 m allows surveying the full Sun at frequencies from 100 MHz to 4 GHz from sunrise to sunset. The 4000 measurements available per second were distributed into 200 channels, yielding a sampling time of 0.1 seconds in each channel. The data were calibrated and cleaned from terrestrial interference using standard routines.

The Nan\c{c}ay Radioheliograph (Kerdraon and Delouis 1997) images the radio emission and was used to locate the position of pulsations. It observes the Sun daily for 7.5 hours centered around 12 UT. The interferometer is operated at discrete frequencies, and we use here data at 237, 327, 410.5, and 432 MHz at a time resolution of 0.15 seconds. For the June event selected, the half-power beam width (major axis of lobe) is 77$''$, 58$''$, 44$''$, and 42$''$, respectively. For the selected event in December, it is 192$''$, 144$''$, 115$''$, and 109$''$, respectively. Comparisons with the VLA have shown agreement on the order of 20$''$ for a typical day (Benz et al. 2005).

The RHESSI satellite is used here for imaging thermal and non-thermal X-rays. It was launched on 5 February 2002, and observes X-rays in the range from 3 keV to 17 MeV with an energy resolution of about 1 keV. Spectral resolution allows differentiating between thermal and non-thermal emissions and selecting the energy range of non-thermal photons. RHESSI rotates to modulate the incoming X-ray flux detected behind the shadowing grids. This allows reconstructing images with high spatial resolution (Hurford et al. 2002). Combined with the high energy resolution, thermal and non-thermal sources can be imaged and identified in different energy ranges (e.g. Emslie et al. 2003; Battaglia and Benz 2006).

SOHO/EIT and GOES/SXI data were used for complementary information about hot plasma. EIT is a normal-incidence, multi-layered mirror instrument (Delaboudini\`ere et al., 1995). It imaged a 7$' \times 7'$ area with a pixel size of 2.62$'' \times 2.62''$. The 195 \AA\ wavelength band was used, including the emission line of Fe XII with diagnostic capabilities for temperatures in the range of 1.1-1.9$\times 10^6$ K. GOES/SXI images the Sun in soft X-rays from 2.1 - 20.7 keV (Hill et al. 2005). The spatial resolution is approximately 10$''$ FWHM.  Images consist of 512 $\times$ 512 pixels with 5$''$ pixel size. GOES/SXI produces full-disk solar images at a 1 minute cadence.

\begin{figure}
\begin{center}
\leavevmode
\mbox{\hspace{0.0cm}\epsfxsize=12cm
\epsffile{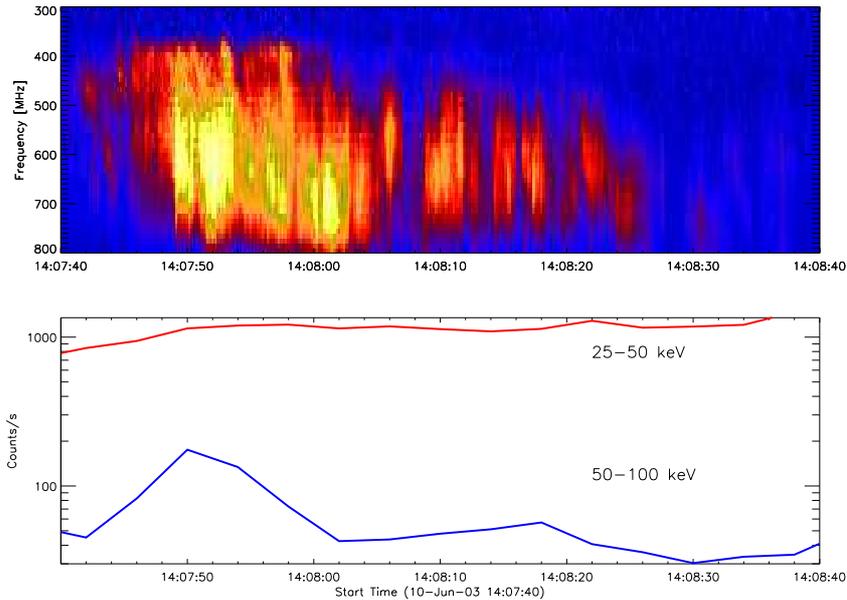}}
\end{center}
%  \vskip-0.5cm
\caption[]{Temporal correlation of radio and X-ray observations of the 10 June 2003 flare. {\it Top:} Extract from Phoenix-2 spectrum showing decimetric pulsations. {\it Bottom:} Simultaneous time profile of hard X-rays observed by RHESSI between 25-50 keV (red) and 50-100 keV (blue).}
  \label{fig:2003_time}
\end{figure}

\subsection{Flare selection}
We selected events from all Phoenix-2 data classified as DCIM and published in Solar and Geophysical Data, starting from RHESSI launch in 2002 until the end of 2007. DCIM stands for 'decimetric events' that differ in spectrograms from burst types I to V at meter waves. DCIM have the subclasses 'pulsations', 'spikes', and 'continuum'. Using Phoenix-2 quicklook images, these subclasses can be distinguished. We followed the selection of Dabrowski and Benz (2009), who found 870 DCIM events and selected those with simultaneous RHESSI observations.  With the requirement of at least 75\% temporal coverage by RHESSI, they found 107 decimetric pulsations.  From this set of simultaneous events they excluded those that were obviously not correlated (delays of more than 20 seconds in cross-correlation). Dabrowski and Benz (2009) finally list 33 pulsations correlating with hard X-rays in detail. Using their list, we selected events that were in the Nan\c{c}ay time window and observable at the Nan\c{c}ay frequencies. In addition we limited the selection to flares that occurred at radial distances larger than 700$''$ from disk center to simplify geometrical interpretation.

Finally, the events of 10 June 2003 14:07-14:09 UT and 5 December 2006 11:15-11:18 UT remained, for which Dabrowski and Benz (2009) report correlation delays relative to X-rays of -2.43 s and 4.07 s, respectively (fitting the cross-correlation coefficient by a Gaussian in delay time). Thus, the events selected here are well time-correlated pulsations at the limb observed by RHESSI.

\begin{figure}
\begin{center}
\leavevmode
\mbox{\hspace{0.0cm}\epsfxsize=12cm
\epsffile{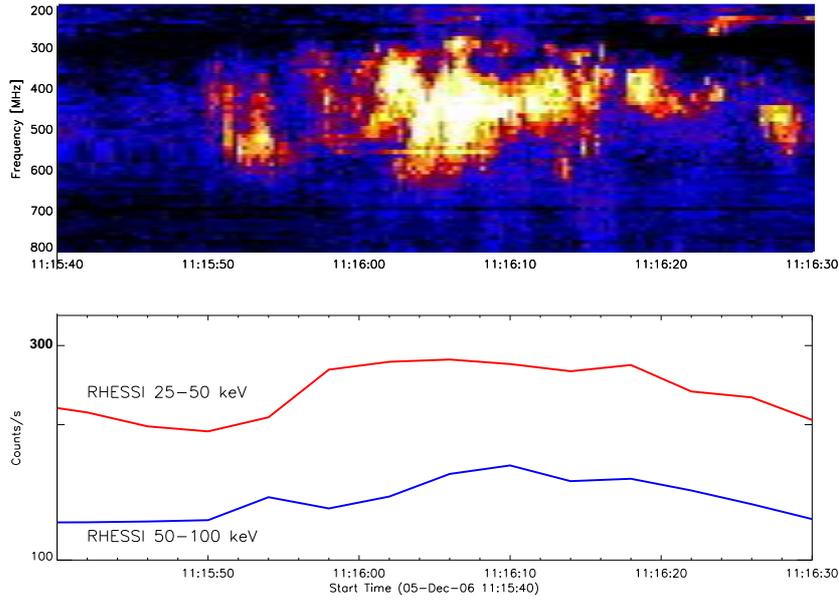}}
\end{center}
%  \vskip-0.5cm
\caption[]{Temporal correlation of radio and X-ray observations of the 5 December 2006 flare. {\it Top:} Extract from Phoenix-2 spectrum showing decimetric pulsations. The observed flux density is logarithmically compressed and indicated by colors from blue to white. {\it Bottom:} Simultaneous time profile of hard X-rays observed by RHESSI between 25-50 keV (red) and 50-100 keV (blue).}
  \label{fig:2006_time}
\end{figure}

The data are presented in Figs. 1 and 2 in spectrum and time. The X-ray time bins are 4 s, about the RHESSI satellite rotation period.

\subsection{Data analysis}
RHESSI X-ray images were constructed for 1 minute time intervals for four energy bands in the thermal and non-thermal range (10-14, 14-18, 18-25, 25-50 keV). The image integrations started at 14:07 UT for the 10 June 2003 event and at 11:15 UT for the 5 December 2006 event. Spectral analysis was possible in the photon energy range 6-40 keV. The low energies are well fitted by a single temperature thermal component. The higher energies can be fitted by a power-law or a second, much hotter thermal component. Evidence from larger flares suggests the non-thermal interpretation. The spectral and spatial analysis indicates thermal emission by the coronal source for the 10-14 keV  range and non-thermal footpoint emission in the 25-50 keV band, while the 14-18 keV and 18-25 keV bands might include some non-thermal emission from the coronal source.

In the 10 June 2003 flare an EIT image taken at 14:12:10 UT and including 12.56 s of integration time was available for complimentary information about the thermal emission at lower temperatures. For the 5 December 2006 flare a GOES/SXI image taken at 11:16:49 UT with 3s integration was used.

The source parameters in Nan\c{c}ay Radioheliograph images were determined at each frequency and in each time bin, fitting automatically a two-dimensional Gaussian as described by Battaglia and Benz (2009). The relevant results are peak location in solar coordinates and peak flux. Solar radio images are generally composed of a practically constant background and a temporary burst component. Only peak fluxes above average flux in the interval are attributed to the pulsating emission and used in the following for plots of individual positions and averages. The standard deviation of the corresponding positions yields an upper limit for the accuracy of the fitted positions (error bars in Figs. \ref{fig:2003_image} and \ref{fig:2006_image}).

\section{Results}

\begin{figure}
\begin{center}
\leavevmode
\mbox{\hspace{0.0cm}\epsfxsize=12cm
\epsffile{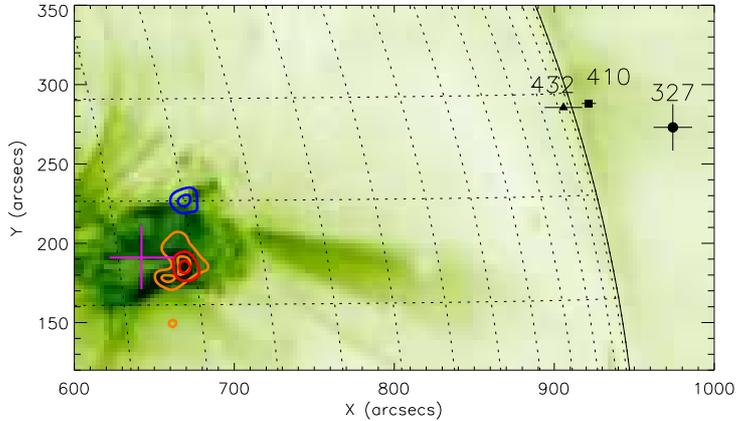}}
\end{center}
%  \vskip-0.5cm
\caption[]{Centroid positions of the radio pulsations in the 10 June 2003 flare measured by the Nan\c{c}ay Radioheliograph overlaid over an EIT image at 195 \AA. The time-averaged centroid positions are shown with error bars indicating the full width at half-power of the distribution of the individual measurements. The observed frequencies in MHz are indicated. The contours indicate RHESSI observations at 10-14 keV (red, thermal) and 18-25 keV (dark yellow, mostly thermal) of the coronal source. The origin of the 25-50 keV (blue, non-thermal) emission is at a different location. The cross marks the H$\alpha$ centroid position and positional accuracy reported by the Kanzelh\"ohe Observatorium (courtesy A. Veronig). The solar coordinate grid and the photospheric limb are indicated for orientation.}
  \label{fig:2003_image}
\end{figure}

Figure \ref{fig:2003_image} shows centroid positions of enhanced radio flux for the 10 June 2003 event. They differ considerably between the different frequencies. The average altitude clearly increases with decreasing frequency. The 432 and 410 MHz average positions are approximately aligned with the direction to the active region, visible in the EUV image, and preferentially with the hardest X-ray source (blue).

The radio sources of the 10 June 2003 flare originated 250$''$-320$''$ (or at least 0.3 solar radii) above the X-ray emission of the coronal source. In view of this large separation, the connection between X-ray and radio sources is not clear. Considering the close temporal correlation between X-ray and radio emission, this is a surprising result.

The EIT image indicates a complex active region with possibly two loop systems, where the southern footpoint of the northern system is close to the northern footpoint of the southern system. For this event, H$\alpha$ observations from Kanzelh\"ohe were available. The two EIT footpoints in the middle coincide with the H$\alpha$ peak position within the positional accuracy of the H$\alpha$ peak, indicated by the error bar. This common footpoint is not visible in X-rays. The total X-ray spectrum can be fitted by a purely thermal electron distribution. The prominent X-ray source at low energies is associated with the southern loop system. It is apparently a coronal source. The harder X-ray source (blue) is a hot spot and possibly also a coronal source related to the northern loop system. The EIT image shows a protrusion from this spot approximately in the direction to the radio sources.

\begin{figure}
\begin{center}
\leavevmode
\mbox{\hspace{0.0cm}\epsfxsize=12cm
\epsffile{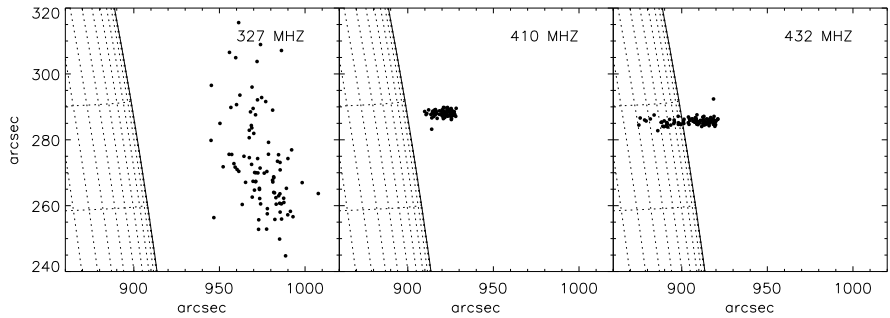}}
\end{center}
%  \vskip-0.5cm
\caption[]{Positions of individual radio pulsation centroids in the 10 June 2003 flare measured by the Nan\c{c}ay Radioheliograph. Only the measurements with flux above the average peak flux are shown. The solar coordinate grid and the photospheric limb are indicated for orientation. The coronal X-ray source(s) (see Figure \ref{fig:2003_image}) are far off the maps to the lower left.}
  \label{fig:2003_scatter}
\end{figure}

 Figure \ref{fig:2003_scatter} presents the individual radio positions at 327, 410, and 432 MHz of time bins with flux above average, thus above background. The centroid positions at 437 and 410 MHz are scattered along a line in $x$-direction. The flare positions in X-rays are far off the picture to the lower left. The angle of the line from the radio positions to the harder X-ray source (blue in Figure \ref{fig:2003_image}) and the $x$-direction is 28$^\circ \pm 2^\circ$. The angle to the softer source (red in Figure \ref{fig:2003_image}) is 37$^\circ \pm 2^\circ$. The time evolution of this scatter was studied, but no systematic motion was found. The 327 MHz positions are different. They scatter parallel to the limb, but the pulsations are very weak or absent (Figure \ref{fig:2003_time}). This lowest frequency may not be relevant for the pulsation as suggested also by other evidence presented later.

\begin{figure}
\begin{center}
\leavevmode
\mbox{\hspace{0.0cm}\epsfxsize=12cm
\epsffile{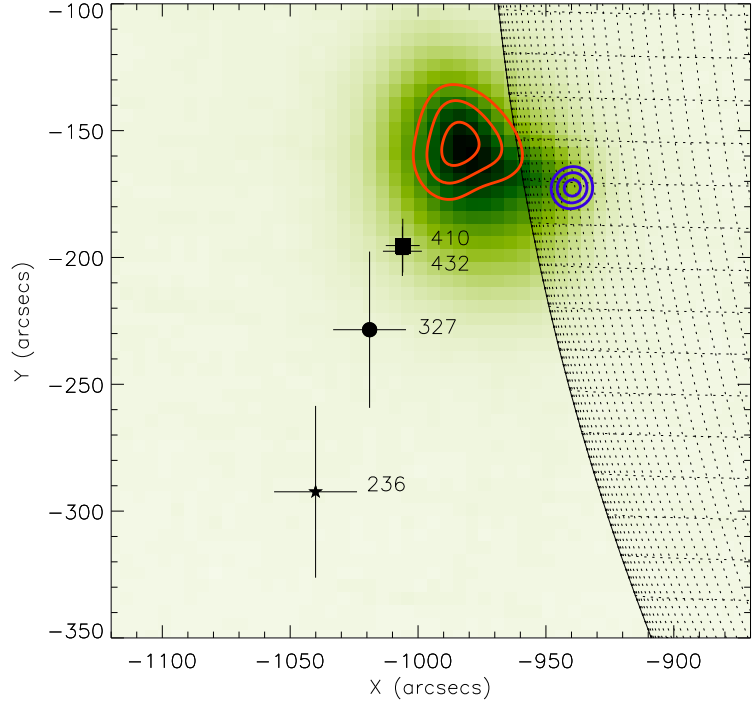}}
\end{center}
%  \vskip-0.5cm
\caption[]{Spatial information on the 5 December 2006 flare. The averaged centroid positions observed by the Nan\c{c}ay Radioheliograph are shown on a GOES/SXT image. The position of the second, stronger interval is displayed for 236 MHz. The contours show RHESSI observations at 18-25 keV (red, coronal source), and 25-50 keV (blue, non-thermal, footpoint source).}
    \label{fig:2006_image}
\end{figure}

\begin{figure}
\begin{center}
\leavevmode
\mbox{\hspace{-1.0cm}\epsfxsize=13cm
\epsffile{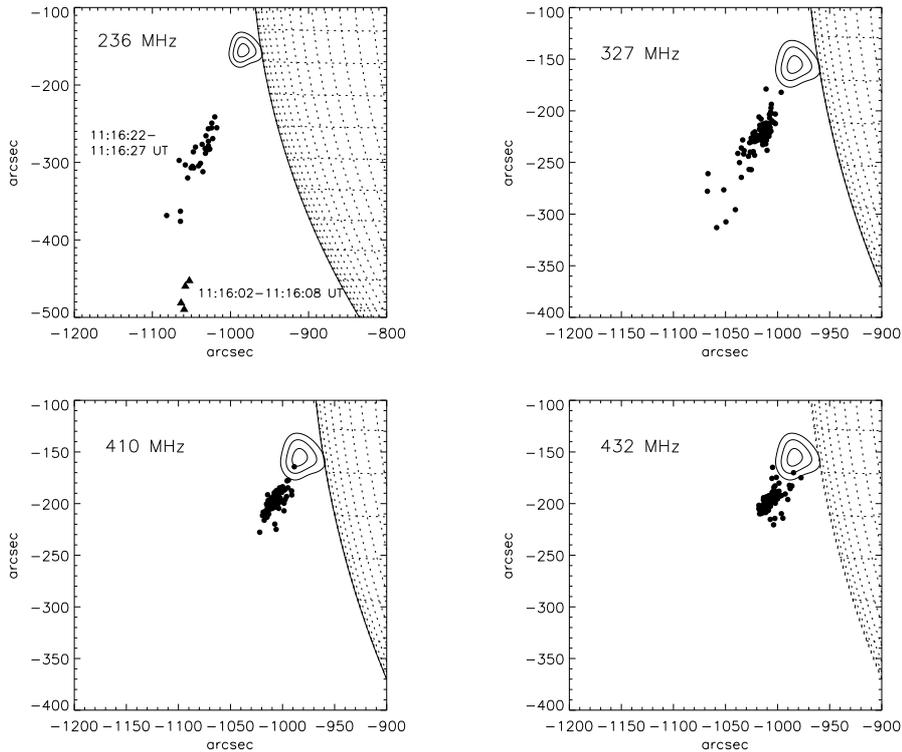}}
\end{center}
%  \vskip-0.5cm
\caption[]{Positions of individual radio pulsation centroids in the 5 December 2006 flare measured by the Nan\c{c}ay Radioheliograph. Only the measurements with flux above the average peak flux are shown. Contours at 50\% 70\% and 90\% of RHESSI Clean images in the 18-25 keV energy band (same as yellow curves in Figure 5) indicate the position and extent of the coronal X-ray source. Note the different scale in the 236 MHz picture.}
  \label{fig:2006_scatter}
\end{figure}

The geometrical situation is even more suggestive in the 5 December 2006 flare. The line of the centroid positions points straight to the coronal X-ray source (red in Figure \ref{fig:2006_image}). The highest frequency source, at 432 MHz, is still 40$''$ (thus 30000 km in projection)  above the peak of the soft X-ray emission. Emission at 236 MHz occurs in two intervals, weak at 11:16:02-08 UT and stronger at 11:16:22-27 UT (Figure \ref{fig:2006_time}). It originates from different locations. Only the second interval is shown in Figure \ref{fig:2006_image}. The first interval is emitted at about twice the distance (see Figure \ref{fig:2006_scatter}). We have investigated the Phoenix-2 spectrogram and polarigram in detail concerning classification of the 236 MHz emissions. The two intervals were found consistent with being part of the pulsations at higher frequencies. There is no indication in spectrum or polarization for another classification, such as Type I, II, or III. No apparent motion is detected at the other frequencies. The X-ray image shows only one 25-50 keV footpoint. The soft X-ray image is consistent with a limb crossing loop and with the hypothesis that the other footpoint is occulted. This suggests that projection effects are minimal.

In the 5 December 2006 flare (Figure \ref{fig:2006_image}), the average projected distance $s$ measured from the center of the coronal X-ray source to the radio source can be fitted by the linear relation

\begin{equation}
s\ \approx\ -386\ \nu\ +\ 1.96 \times 10^5 \ \ {\rm [km]},
\end{equation}
where $\nu$ is the observing frequency in MHz. If Eq. (1) is extrapolated, it suggests that the position of 510$\pm$10 MHz emission coincides with the peak of the coronal source ($s=0$). This frequency is within the range over which pulsations are observed in the spectrogram (Figure \ref{fig:2006_time}). We may add here that the plasma frequency derived from the X-ray emission measure and source volume amounts to 560 MHz. A similar extrapolation can be made for the 10 June 2003 flare, although the observed radio sources are much farther away from the X-ray sources. The result is the same: Extrapolating the 410 MHz and 432 MHz positions, the location of the highest observed frequency at which pulsations are observed, 830 MHz, would be in or near the coronal X-ray source (blue in Figure \ref{fig:2003_image}). The scaling factor in Eq. (1) is a factor of 5 larger for the 10 June 2003 event.

Figure \ref{fig:2006_scatter} shows that the individual centroid positions are scattered along the same line for all frequencies. Again, the lower the frequency, the higher in the corona the radio emission originates. The 236 MHz emission needs a special discussion. During the first interval, 11:16:02-08 UT, the radio emission was observed at a projected distance of 360$''$ from the coronal X-ray source and shifted to the right of the line from the coronal X-ray sources to the other radio sources. The radio emission during the second time interval, 11:16:22-27 UT, was much stronger, originated closer to the X-ray source and straight on the above line. The centroid positions at 236 MHz do not scatter excessively and appear to be related to the pulsations at higher frequency.

%\vskip-5cm
\begin{figure}
\begin{center}
\leavevmode
\mbox{\hspace{0.0cm}\epsfxsize=12cm
\epsffile{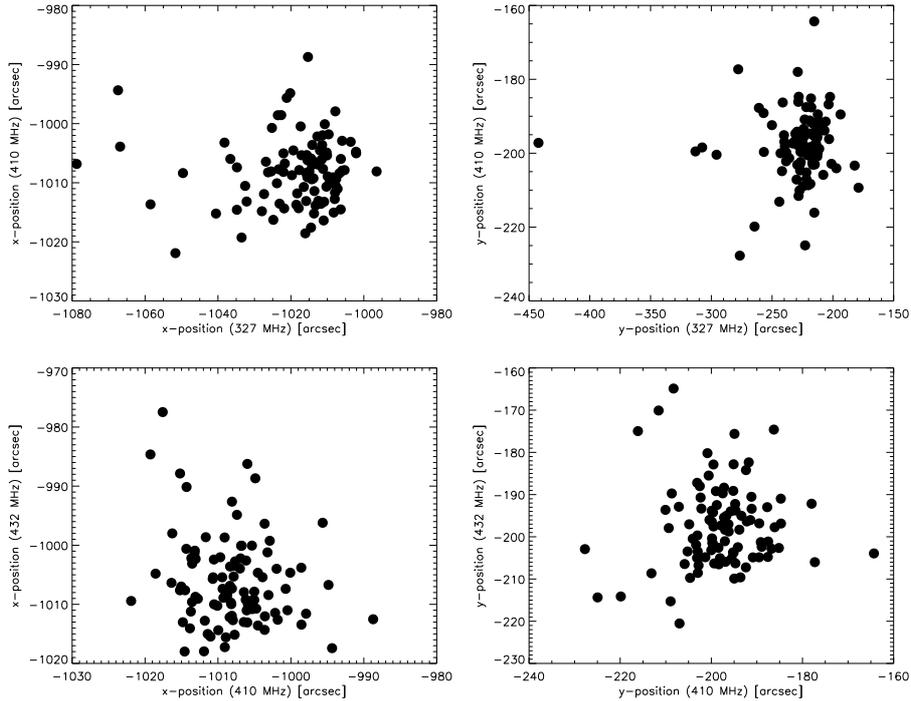}}
\end{center}
%  \vskip-0.5cm
\caption[]{Comparison of simultaneous x-positions (left) and y-positions (right) at different frequencies for the 5 December 2006 flare. Only fluxes above average in both frequencies are shown.}
  \label{fig:scatter_correlation}
\end{figure}

\section{Discussion}

The time-correlating radio and X-ray sources are clearly separated in space. The average radio positions are at higher altitude, increasing to lower frequency. Nevertheless, a remarkable relation between the two emissions was found. In the event of 5 December 2006 (Figure \ref{fig:2006_image}), the average positions at different frequencies are on a line pointing straight toward the coronal X-ray source. They are at larger apparent distance in the 10 June 2003 event, and point approximately to the X-ray source at higher X-ray energies (Figure \ref{fig:2003_image}).

Even more surprising, the scatter of individual centroid positions is not random in both flares. The individual positions are preferentially aligned out on the above line in the direction to the coronal X-ray flare site in the 5 December 2006 (Figure \ref{fig:2006_scatter}). For the 10 June 2003 event, the alignment is at an angle of some 30$^\circ$ from this line (Figure \ref{fig:2003_scatter}).

\begin{figure}
\begin{center}
\leavevmode
\mbox{\hspace{0.0cm}\epsfxsize=12cm
\epsffile{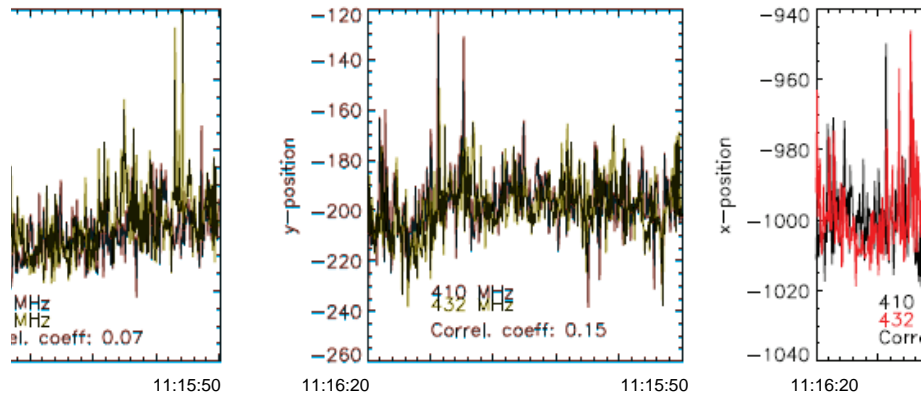}}
\end{center}
%  \vskip-0.5cm
\caption[]{Comparison of positions in adjacent frequencies for the 5 December 2006 flare. All time bins are shown.}
  \label{fig:pulstime}
\end{figure}

Scattering of a source position at a given frequency may have several reasons. First, we discuss the possibility of ionospheric scattering. Variable refraction would move the positions of nearby frequencies, measured simultaneously, in the same direction. Thus, the excursions in both $x$-direction and $y$-direction would be similar, along a line, and correlated for adjacent frequencies. No correlation between the positions at different frequencies is observed (Figure \ref{fig:scatter_correlation}). A plot of the $x$-positions and $y$-positions  vs time, however, reveals occasional correlation of the baselines at 410 and 432 MHz (Figure \ref{fig:pulstime}). Thus, the relatively low cross-correlation value and the scatter in Figure \ref{fig:scatter_correlation} result from uncorrelated excursions, which cannot originate from ionospheric scattering. The scatter of the 10 June 2003 radio positions has the same property.

A second cause for linear scatter could be a seesaw of the centroid between two positions. The positions of the bins below and above average flux density were compared. Their distributions in $x$ and $y$ are similar. The average positions of low and high flux are the same in both events and at all frequencies. Thus there seems to be only one source moving randomly in position.

In the absence of contrary evidence, we suggest that the linear spreads of the centroid positions in the 5 December 2006 event in all four frequencies and in the 10 June 2003 event at 410 and 432 MHz are real. This is supported by Figure \ref{fig:2006_image} showing a linear alignment of the average sources in decreasing frequency in line with the coronal X-ray source. Figure \ref{fig:2003_image} indicates the same behavior for 432 and 410 MHz. Only the scatter of the 327 MHz positions on 10 June 2003 deviates from this general behavior. As noted before, its flux is very low such that the emission cannot be classified in the spectrogram (Figure \ref{fig:2003_time}). Throughout the day, the Nan\c{c}ay Radioheliograph has observed a low intensity noise storm (importance I) at 327 MHz reported at +955$''\pm 10''$ W, 246$''\pm 10''$ N (Solar-Geophysical Data, 2003). This position is only 2 standard deviations southeast of the average position measured here. Thus the 327 MHz source is most likely not pulsating emission.

The scatter in pulsation position appears to have two components: ({\it i}) rapid fluctuations that are not correlated in frequency even as close as 410 and 423 MHz, and ({\it ii}) correlated changes in positions of less than $\pm 10''$ that appear as slow variations of the baseline in time (Figure \ref{fig:pulstime}). The latter seems to be caused by a systematic motion of the pulsations' centroid at both frequencies.

\section{Conclusions}

The positions of radio pulsations and soft and hard X-ray sources are found far apart despite the good correlation in time. The projected difference is at least 30000 km. This is far beyond the distance an Alfv\'en wave would travel within the delay time measured by cross-correlation. It is therefore likely that the trigger signal is propagated by energetic particles.

The radio emission originates consistently above the coronal X-ray source. Evoking the loss-cone instability of electrons trapped in such loops would predict the radio sources near the mirroring magnetic fields, thus near the footpoints of the loops and below the coronal X-ray region. This clearly contradicts the old scenario proposing decimetric pulsations to be caused by electrons trapped in flare loops (see Introduction).

We cannot distinguish between the two remaining scenarios of the cause of the pulsed modulation:  pulsating acceleration or particles accelerated elsewhere radiating in an oscillating MHD structure. As the radio sources are displaced from the main acceleration site (generally assumed to be close to the coronal X-ray source), the pulsating acceleration scenario would suggest additional energy release, although less energetic, in a larger volume including the radio sources. This can be envisaged as a current sheet extending upward from the flaring loops possibly behind an ejected plasmoid (see Introduction). We have not been able to find ancillary observations for the two selected events from coronographs, or other X-ray or EUV instruments to confirm this hypothesis. If true, decimetric pulsations would trace the coronal current sheet.

The comparison of positions indicates that the radio sources at different frequencies are aligned on a structure pointing upward from the coronal X-ray source in both flares. The scatter of positions at a given frequency is along the same line in one case, and at an angle of some 30$^\circ$ in the other case. This suggests that the radio emission is emitted at a frequency related to the local density or magnetic field (or both) which decrease with altitude. The scatter at a given frequency then could be the result of the source occurring on individual field lines with different density or magnetic field.

In the December 5 2006 event, emission was also observed at 236 MHz. This source moved 220$''$ inward within 20 s (Figure \ref{fig:2003_scatter}), thus exceeding the commonly assumed values of the Alfv\'en velocity by more than an order of magnitude. The motion indicates that the conditions (density and/or magnetic field) at the first source were the same as 20 s later in the second source. Most likely, the emission site moved from a relatively dense region high in the corona to a lower region having the same density. At the higher frequencies and closer to the coronal X-ray source, no motion was observed within the accuracy of the observation.

Due to instrumental limitation, radio pulsations at low frequencies were selected. In addition, the positional information refers to the lower end of the frequency range of the selected events. We cannot exclude that at higher decimetric frequencies radio pulsations occur near or in the coronal X-ray source, as suggested by Eq.(1). The confirmation of such a hypothesis would require a new radio interferometer at frequencies above 500 MHz.

\begin{acknowledgements}
We thank Christian Monstein and Andreas James for constantly improving and operating the Bleien radio spectrometers. The construction of the spectrometers was financially supported by the Swiss National Science Foundation (grants 20-113556 and 200020-121676). The NRH is funded by the French Ministry of Education, the CNES and the R{\'e}gion Centre in France.

\end{acknowledgements}

%------------------------------------------------------------------------------
%BIBLIOGRAPHY
%------------------------------------------------------------------------------
{}

\end{article}
\end{document}